\documentclass[12pt]{article}

\usepackage{amsmath}
\usepackage{amssymb}
\usepackage{eufrak}

\newcommand{\eps}{\varepsilon}
\newcommand{\ga}{\gamma}

\begin{document}

\title{Cascade emission of $\gamma$-quanta by highly excited nuclei
\footnote{Zh. Eksp. Teor. Fiz. {\bf 65}, 12--23 (1973) [Sov. Phys. JETP
{\bf 38,} No.~1, 6-11 (1974)]}}

\author{V.G. Nosov$^{\dagger}$ and A.M. Kamchatnov$^{\ddagger}$\\
$^{\dagger}${\small\it Russian Research Center Kurchatov Institute, pl. Kurchatova 1, Moscow,
123182 Russia}\\
$^{\ddagger}${\small\it Institute of Spectroscopy, Russian Academy of Sciences, Troitsk, Moscow Region,
142190 Russia}
}

\maketitle

\begin{abstract}
The thermodynamics of the electromagnetic radiation from heated nuclei is developed on
basis of the Landau theory of a Fermi liquid [1]. The case of non-spherical nuclei is
considered, in which the quasiparticle energy spectrum is not distorted by the residual
interactions that affect the thermodynamic behavior of the system. The
number of quanta per cascade and mean-square fluctuation are calculated; the
$\gamma$-quantum spectrum of the whole cascade is also obtained. The formulae can be
used to determine the entropy and temperature of the initial nucleus by various methods.
The effective nucleon (quasiparticle) mass in nuclear matter is determined by comparison
with the experimental data. The region of validity of the theory and some possibilities
of its extension on the basis of new experiments are discussed.
\end{abstract}

\section{Introduction}

Radiative transitions between the lowest levels of nuclei exhibit a great diversity
in their intensities and multipolarities. In addition to the usually intense transitions
between the ``collective" (rotational and vibrational) levels, there are encountered
also isomeric $\gamma$ transitions, the high degree of hindrance of which may be due to the
large change in the nuclear spin and also to other causes. Even a schematic tentative
classification and a very cursory discussion of the particular interest that may attach
to any particular modification of the radiative transitions between any two concrete
energy levels of the nucleus would greatly exceed the scope of the present article.
In nuclei that are not too light, however, the number of levels that lend themselves
to transitions with $\gamma$-quantum emission increases sharply, with increasing
excitation energy, and the individual features then become gradually averaged. A
well known example may be the observed spectra of $\gamma$ rays of radiative capture
of thermal neutrons [1]. It is clear that for a sufficiently heavy nucleus, the overall
picture of such a phenomenon characterizes not the individual levels, but more
readily a certain region of the energy spectrum of the nucleus as a whole. It is
natural to assume that rather general laws of phenomenological  character come to
the forefront here. Unification of the dominant mechanism of the process becomes
manifest, e.g., in the fact that these electromagnetic radiation spectra [2] of many
different nuclei exhibit great similarities.

Incidently, at the excitation energy $\sim 8$ MeV, which we referred to in this case,
certain striking qualitative differences still remain. Since they are due to phenomena
that are of very great importance in nuclear physics, we shall consider
the examples that illustrate this fact in somewhat greater detail. We consider the
situations on both sides of osmium  ($Z = 76$), where the spectra of  $\gamma$
quanta [2] have been investigated experimentally in sufficient detail. Many nuclei
of the chemical elements preceding osmium have spectra of rather standard form,
namely the energy distribution of the quanta has a maximum at $\eps\leq 2$ MeV,
after which it drops off rapidly towards the limiting value $\eps = E$ (the excitation
energy of the nucleus in the initial state). However, this ``temperature" maximum
becomes considerably smoothed out even for the first element following osmium, namely
iridium ($Z = 77$), and the transitions in the hard part of the $\ga$ spectrum are
simultaneously smoothed out. This characteristic deformation of the
spectrum of quanta is further developed in the case of $_{78}$Pt, where, roughly
speaking, the areas under both maxima become comparable. The emission spectra of
$_{79}$Au reveals in practice only one maximum adjacent to the hard edge $\eps=\eps_n$
($\eps_n$ is the neutron binding energy). In the elements that follow, $_{80}$Hg
and $_{81}$Tl, the relative area under the hard part of the spectrum continues
to increase. It is curious to note that this feature becomes much more sharply
pronounced in the case of the doubly magic nucleus $_{82}$Pb$_{126}^{208}$. When the
preceding isotope Pb$^{207}$ captures a thermal neutron, practically 100\% of all
the radiative transitions go directly to the ground state; in other words, the
spectrum of the cascade degenerates into a single line  $\eps=\eps_n$.

The picture outlined above is probably brought by about a unique phenomenon that
becomes manifest not only in the spectra of the $\ga$-quantum cascades; the
equilibrium shape of the nucleus also changes in the immediate vicinity of osmium.
The properties of the energy spectrum of spherical nuclei located in the region
adjacent to the doubly-magic nucleus are strongly influenced by the residual
interaction between the quasiparticles\footnote{This residual interaction has a
macroscopic structure that influences the behavior of the nucleus as a whole.
It was analyzed in [3].}. To the contrary, on the other side
of the phase transition point [4], such an influence apparently ceases to be
decisive in any manner. The non-spherical shape of the nuclei in this region
is quite natural, for when there is no interaction whatever between the particles
the instability of the spherical configuration is proved by direct calculation.

We consider below the properties of electromagnetic radiation of all the nuclei
pertaining to this non-spherical ``normal" phase (see [4]).

\section{Thermodynamics of electromagnetic radiation of
excited nuclei}

According to Fermi-liquid theory [1,6], the behavior of this liquid is determined by
quasiparticles that obey the Pauli principle and are sufficiently close to the Fermi
boundary. In the state of thermal equilibrium, the usual Fermi distribution holds
\footnote{We call attention to the connection between the condition $T \ll \eps_n$
($-\eps_n$ is the chemical potential) for the applicability of the theory,
on the one hand, and the very possible existence of a compound nucleus,
on the other. In the opposite case, the neutrons having a binding energy $\eps_n$
would be emitted from the nucleus ``instantaneously," bypassing the stage of
establishment of thermal equilibrium.}
\begin{equation}\label{1}
    n(\eps')=\frac1{e^{\eps'/T}+1},
\end{equation}
where $\eps'$ is the energy of the quasiparticle reckoned from the chemical potential,
and $T$ is the temperature. We now explain the predominant mechanism of the process.
In accordance with the accepted concepts, the quasiparticles move freely inside the
nuclear matter. However, when they strike the transition region on the nuclear surface,
they are reflected from the latter, i.e., they are accelerated. This makes the emission
of electromagnetic quanta possible\footnote{We mention also another possibility:
radiation could be produced also when quasiparticles collide with one another.
However, the number of such collisions is proportional to the cube of the
temperature, whereas the effect calculated by us is linear in the temperature
(see formula (2) below). Therefore the mechanism of collisions of individual
quasiparticles with the ``wall" of the nucleus should be regarded as predominant;
see also footnote 2.}.

On the average, we ascribe radiation to an individual quasiparticle in accordance
with the law $f(\eps)d\eps$, where $\eps$ is the energy of the $\ga$ quantum
(the form of the function $f(\eps)$ will be established below). To go over to
the true probability distribution $w(\eps)d\eps$, we must take into account the
entire aggregate (1) of the quasiparticles obeying the Pauli principle.
Actually this reduces to multiplication by the product
$n(\eps')[1 - n(\eps' - \eps)]$ and integration over the fermion energies.
Simple integration yields
\begin{equation}\label{2}
    \int_{-\infty}^\infty n(\eps')[1 - n(\eps' - \eps)]d\eps'=\frac\eps{e^{\eps/T}-1},
    \quad w(\eps)d\eps=\frac{f(\eps)\eps d\eps}{e^{\eps/T}-1}.
\end{equation}
We now consider the question from a somewhat different point of view. The radiation
wavelength is long relative to the dimensions of the nucleus and is principally of
the electric-dipole type. The average level density of the system changes
significantly only over energy intervals of the order of the temperature.
In other words, when the excitation energy changes by an amount $\eps\ll T$,
the energy characteristics of the spectrum of the nucleus as a whole, averaged
over many quantum states, remain practically constant. Therefore in the limit as
$\eps\to 0$ there remains only the cubic dependence $w(\eps) \propto\eps^3$
of the probability of the process on the transition energy, a dependence
characteristic of dipole emission. Taking (2) into account we therefore have
\footnote{It is well known that owing to recoil of the nucleus as a whole, which has a
charge $Ze$, the neutron component is also capable of electric dipole emission
(see, e.g., [7]). In fact therefore, both proton and neutron quasiparticles
take part in the process in question. However, it is clear from the character
of the presented results that the form of the required expressions is determined
only by the temperature of the nucleus and explicit allowance for the two-component
character of the nuclear matter would not influence them. We therefore start for
the time being from a simplified picture of quasiparticles of one type, the
statistical distribution of which is given by formula (1).}
\begin{equation}\label{3}
    f(\eps)=\mathrm{const}\cdot\eps^3.
\end{equation}
For our purposes there is no need to calculate the absolute value of the probability
$\Gamma_\ga/\hbar$ of the radiation per unit time ($\Gamma_\ga$ is the radiative width).
We can deduce even from (2) and (3), however, how this quantity depends on the
temperature of the nucleus (or on the excitation energy; see (15)). Since the
constant factor in the right-hand side of (3) depends neither on $\eps$ nor on $T$,
the integration of the second formula of (2) yields
\begin{equation}\label{4}
    \Gamma_\ga\propto T^5.
\end{equation}
The $\ga$-quantum energy distribution $w(\eps)d\eps$ will now be renormalized to a
unit total probability of its radiation
\begin{equation}\label{5}
    w(\eps)d\eps=\frac1{24\zeta(5)T^5}\frac{\eps^4d\eps}{e^{\eps/T}-1},\quad
    \int_0^\infty w(\eps)d\eps=1.
\end{equation}
Here
$$
\zeta(s)=\frac1{\Gamma(z)}\int_0^\infty\frac{x^{s-1}dx}{e^x-1}=
\sum_{n=1}^\infty\frac1{n^s}
$$
is the Riemann $\zeta$ function. We need also the average energy $\bar{\eps}(T)$
of the quantum emitted by the nucleus with temperature $T$,
\begin{equation}\label{6}
    \bar{\eps}=\int_0^\infty\eps w(\eps)d\eps=
    \frac{\pi^6}{189}\frac{T}{\zeta(5)}\simeq 4.91 T.
\end{equation}

As the nucleus radiates, it becomes cooler; the running values of the excitation
energy and of the temperature will be denoted by $E'$ and $T'$, respectively.
For the average number $\nu$ of the quanta in the cascade we obtain
\begin{equation}\label{7}
    \nu=\int_0^E\frac{dE'}{\bar{\eps}(T')}=\frac{189\zeta(5)}{\pi^6}
    \int_0^E\frac{dE'}{T'}\simeq\frac{S}{4.91},
\end{equation}
where $E$ is the energy of the initial state of the nucleus and $S$ is its entropy.
Relation (7) makes it possible to determine the latter from experiment.

How are the probability distributions (5) of different quanta of the cascade
interrelated? Since the function $w(\eps, T)$ depends on $T$ as a parameter,
its form is determined by the prior history and by the total energy of the
entire preceding radiation that caused the nucleus to cool down to the
temperature in question. However, reasoning somewhat formally, such a relation
between the $\ga$ quanta of the cascade is due to the fact that each of them is
characterized by an energy $\eps$. We now change over to another variable $s$,
which is the entropy carried away by the quantum from the nucleus. Obviously,
$s = \eps/T$, and the distribution (5) can be rewritten in the form
\begin{equation}\label{8}
    w(s)ds=\frac1{24\zeta(5)}\frac{s^4ds}{e^s-1}
\end{equation}
Thus, in terms of $s$ the $\ga$ quanta are statistically independent. This
simplifies very greatly the calculation of the fluctuations of the number of
quanta in the cascade. We write down the corresponding mean values
\begin{equation}\label{9}
\begin{split}
    \overline{s}=&\int_0^\infty sw(s)ds=\frac{\pi^6}{189\zeta(5)}\simeq 4.91,\quad
    \overline{s^2}=\int_0^\infty s^2w(s)ds=30\frac{\zeta(7)}{\zeta(5)}\simeq 29.2,
    \\ &\overline{(\Delta s)^2}=\overline{s^2}-\overline{s}^2\simeq 5.11
    \end{split}
\end{equation}
(the first of these formulas, in fact, is a restatement of (6) in terms of other units).

We consider next a portion of the cascade consisting of $\nu\,'$ successively
emitted quanta. For the fluctuation of the entropy $S'$ pertaining to this section
we have the expressions
\begin{equation}\label{10}
    \overline{(\Delta S')^2}={\nu}\,'\overline{(\Delta s)^2}\simeq 5.11\nu\,',\quad
    \overline{\left(\Delta\frac{S'}{\nu\,'}\right)^2}\simeq\frac{5.11}{\nu\,'}.
\end{equation}
The last of the formulae (10) determines the fluctuation of the entropy $S'/\nu\,'$
per $\ga$ quantum. The same quantity admits also of another definition: the
considered section can be determined by specifying the constant $S'$, and the
number of quanta needed to produce this entropy drop can be regarded as fluctuating.
Therefore
\begin{equation}\label{11}
    \overline{\left(\Delta\frac{S'}{\nu\,'}\right)^2}=\frac{{S'}^2}{{\nu\,'}^2}
    \overline{(\Delta\nu\,')^2}.
\end{equation}
Combining (11) with (10) and taking (7) into account, we extend the final formula
to include the entire cascade:
\begin{equation}\label{12}
    \overline{(\Delta\nu)^2}=\left\{30\frac{\zeta(7)}{\zeta(5)}-
    \left[\frac{\pi^6}{189\zeta(5)}\right]^2\right\}\frac{\nu^3}{S^2}
    \simeq 0.0433 S.
\end{equation}
Here we have one other method of measuring the entropy of the initial nucleus,
but this time from fluctuations of the number of the $\ga$ quanta per cascade
(cf. (7)). Eliminating $S$ from (7) and (12), we obtain the relation
\begin{equation}\label{13}
    \sqrt{\overline{(\Delta\nu)^2}}\simeq 0.461\sqrt{\nu},
\end{equation}
which may turn out to be useful to verify the mechanism of the process.

The energy spectrum $W(\eps)d\eps$ for the $\ga$ quanta of the entire cascade
is made up of distributions of the type (5) for each of them. Taking also (6)
into account, we have
\begin{equation}\label{14}
    W(\eps)d\eps=d\eps\int_0^E w(\eps,T')\frac{dE'}{\eps(T')}=
    d\eps \eps^4\frac{63}{8\pi^6}\int_0^T\frac{C(T')dT'}{{T'}^6(e^{\eps/T'}-1)}
\end{equation}
where $C(T) = dE/dT$ is the specific heat of the nucleus. Its dependence on the
temperature is determined by the formulas
\begin{equation}\label{15}
    E=\tfrac12aT^2,\quad C=S=aT=\sqrt{2aE},
\end{equation}
which follows from the Fermi-liquid theory [1,6] (see also [8], where rather weighty
arguments were first advanced favoring such an equation of state of the nucleus).
Substituting (15) in (14) and introducing the integration variable $x = \eps/T'$,
we obtain ultimately
\begin{equation}\label{16}
    W(\eps)d\eps=d\eps\frac{63}{8\pi^6}a\int_{\eps/T}^\infty\frac{x^3dx}{e^x-1},\quad
    \int_0^\infty W(\eps)d\eps=\nu
\end{equation}
(in practice it frequently turns out to be convenient to express the coefficient $a$
in terms of the thermodynamic quantities of the initial state of the nucleus in
accordance with (15)). The integral in this formula
\begin{equation}\label{17}
    J(y)=\int_{y}^\infty\frac{x^3dx}{e^x-1}=\frac{\pi^4}{15}-\frac{y^3}3D(y)
\end{equation}
can be easily investigated. It is expressed in terms of the Debye function $D(y)$,
which determines the well-known interpolation for the specific heat of a solid (see,
e.g. [6]).

In addition to the number of quanta $W(\eps)$ per unit change of the variable $\eps$,
we introduce also the distribution of the energy $\mathcal{E}(\eps)d\eps=\eps W(\eps)d\eps$
and the spectrum of the $\ga$ quanta of the cascade:
\begin{equation}\label{18}
    \mathcal{E}(\eps)d\eps=d\eps\frac{63}{8\pi^6}a\eps J\left(\frac\eps{T}\right), \quad
    \int_0^\infty\mathcal{E}(\eps)d\eps=E.
\end{equation}
Taking (16), (7), and (15) into account, we obviously have
\begin{equation}\label{19}
    \overline{\overline{\eps}}=\frac{\int_0^\infty\mathcal{E}(\eps)d\eps}
    {\int_0^\infty W(\eps)d\eps}=\frac{\pi^6}{189\zeta(5)}\frac{E}S=
    \frac{\pi^6}{189\zeta(5)}\frac{T}2\simeq 2.45 T
\end{equation}
for the quantum energy $\overline{\overline{\eps}}$ averaged over the spectrum (cf. (6)).
The function $\mathcal{E}(\eps)$ has a maximum \footnote{To the contrary, the theoretical
expression for $W(\eps)$, determined by formula (16), is a monotonic function. When its
maximum is observed in experiment (see the Introduction above), this is due to the fact that
the theory is not applicable to the softest part of the spectrum. With respect to
the presently available experimental data, the main shortcoming of the proposed
theory is the narrowness of the region of its applicability. The corresponding
criterion will be obtained below, see formulas (25) and (26). The pertinent questions
will be analyzed in greater detail in the concluding sections of the article.}.
To determine its position, we equate the derivative to zero. The transcendental equation
\begin{equation}\label{20}
    \int_0^\infty\frac{x^3dx}{e^x-1}=\frac{y^4}{e^y-1}
\end{equation}
is easy to solve: $y \simeq 2.89T$. Thus,
\begin{equation}\label{21}
    \eps_{max}\simeq 2.89 T.
\end{equation}
This is possibly one of the most convenient methods of determining the temperature of a nucleus.

In view of the importance of the question, we present also a convenient formula that
expresses the initial spectrum $W(\eps)$ directly in terms of $\eps_{max}$:
\begin{equation}\label{22}
    W(\eps)\thickapprox 0.137\frac{E}{(\eps_{max})^2}J\left(\frac{\eps}T\right).
\end{equation}

The limits of applicability of the theory are determined by the requirement
\begin{equation}\label{23}
    \overline{\eps}(T')\ll E'.
\end{equation}
In this sense, the nature of the violations near the hard edge $\eps\ll E'$ of the
spectrum is quite clear. The energy conservation law forbids the emission of quanta
with $\eps > E$, and the theoretical expressions (16) and (18) lead us to nonzero
intensities, although they do prove to be exponentially small. However, in the
softest regions of the spectrum, the theory also ceases to be
valid. Indeed, according to (6) the characteristic energy $\overline{\eps}(T')$
for the $\ga$ quantum is proportional to the temperature of the radiating nucleus, since
its excitation energy $E'$ depends on the temperature quadratically (see (15)),
i.e., it decreases more rapidly.

To make the criterion more precise, it is simplest to take into account the fact that,
according to (6) and (7), we have the proportion
\begin{equation}\label{24}
    \overline{\eps}(T')/T'=S/\nu.
\end{equation}
Replacing the numerator of the left-hand side by the running excitation energy $E'$,
we transform the resultant relation in accordance with (15):
$$
\frac{E'}{T'}=\frac12\sqrt{2aE'}=\frac12\sqrt{\frac{S^2}{E}}\sqrt{E'}=\frac{S}2
\sqrt{\frac{E'}{E}}.
$$
We now equate the result to the right-hand side of (24):
$$
\frac{S}2\sqrt{\frac{E'}{E}}=\frac{S}\nu,\quad E'=4\frac{E}{\nu^2}.
$$

We have obtained here the excitation energy at which $\eps\sim E'$; it determines
the lower limit (with respect to the energy of the emitted quantum) of the
applicability of the theory. Therefore the sought criterion takes the form
\begin{equation}\label{25}
    4E/\nu^2\ll\eps\ll E.
\end{equation}
Consequently, the region of thermal emission of the nucleus exists under the condition
\begin{equation}\label{26}
    \nu^2\gg 4.
\end{equation}

\section{Comparison with experiment. Effective mass of the nucleon (quasiparticle)}

The region of heavy non-spherical nuclei, in which sufficiently systematic experimental
investigations of the $\ga$-cascade spectra were made, extends from samarium-gadolinium
to osmium\footnote{lt is difficult to identify more specifically the chemical element pertaining
to its lower limit, since the phase state of the nucleus is more sensitive here to
the number of neutrons $N$. To the contrary, for the spherical nuclei in the
vicinity of lead, the position of the phase-transformation point depends to a
greater degree on the number of protons. Insofar as can be judged from the
experimental data, osmium ($Z = 76$) is located precisely at the point of transition
of the non-spherical nuclei to spherical ones, or, at any rate, is very close to it;
see also the Introduction.}.
A comparison of the data of [2] with the theory developed in the preceding section was
made for the spectra of radiative capture of thermal neutrons by ten different nuclei.
According to formula (21) (see also the texts pertaining thereto) we determined the
temperature of the initial compound nucleus. Since its excitation energy $E = \eps_n$
is also known, relations (15) enable us to calculate the entropy and the specific heat.
All these results are listed in the table.

The condition (26) may not be well satisfied in the case of capture of thermal neutrons.
From this point of view, an advantageous method of monitoring the obtained temperatures
is a comparison of the theoretical spectra of the $\ga$ quanta with the experimental ones.
The figure shows the measured spectra [2] (in units of quanta/MeV) with those calculated
by formula (22). There is apparently a definite correlation with the total number $\nu$,
calculated in accordance with (7), of the ``evaporation" quanta in the cascade (i.e.,
due to the considered thermal mechanism). To the extent that the area under the
theoretical spectrum $W(\eps)$ approaches $2$, the agreement, generally speaking,
becomes much worse. To the contrary, even at $\nu\approx 4$ satisfactory agreement is
observed in a certain spectral region that does not contradict the criterion (25)
\footnote{We mention in this connection the compound actinide nucleus Th$^{233}$. The following
results were obtained for it: $T = 0.72$ MeV, $E = 4.96$ MeV, $\nu = 2.8$; the agreement
over the spectrum turned out to be poor. The case of the even-even nucleus Sm$^{150}$
is curious in the following respect: it is well known that in accordance with the
spectroscopic data it is spherical in the ground state, but the Curie point lies very
close to its position in the periodic table. In this case, the characteristics of the
$\ga$-quantum cascade (see the table and the figure) do not deviate in any striking
manner from the general picture, and consequently, at $E = 8$ MeV we already have a
non-spherical phase (see also the introduction). It is still difficult to indicate
more accurately the excitation energy at which the phase transition takes place.
To avoid misunderstandings, we note that there is apparently no phase transition
whatever when non-spherical nuclei are excited; see [4,5].}.
On the whole, qualitative considerations suggest that, owing to the excessively small
number of the quanta in the cascade, the method considered here may overestimate
somewhat the nuclear temperature.

\begin{table}
\begin{tabular}{|c|c|c|c|c|c|}
\hline
Compound nucleus & $E$, MeV &$T$, MeV & $S=C$ & $\Delta T,$ MeV & ${m^*}/{m_n}$ \\
\hline
$_{62}$Sm$_{88}^{150}$ & 7.98 &0.80 & 20.0 & 0.18 & 1.01 \\
$_{63}$Eu$_{89}^{152}$ & 6.29 &0.74 & 17.0 & 0.18 & 0.91 \\
$_{64}$Gd$_{92}^{156}$ & 8.53 &0.87 & 19.6 & 0.20 & 0.88 \\
$_{64}$Gd$_{94}^{158}$ & 7.93 &0.90 & 17.6 & 0.21 & 0.75 \\
$_{66}$Dy$_{99}^{165}$ & 5.64 &0.90 & 12.5 & 0.25 & 0.51 \\
$_{68}$Er$_{100}^{168}$ & 7.77 &1.00 & 15.6 & 0.25 & 0.56 \\
$_{72}$Hf$_{106}^{178}$ & 7.62 &0.83 & 18.4 & 0.19 & 0.75 \\
$_{73}$Ta$_{109}^{182}$ & 6.06 &0.66 & 18.4 & 0.15 & 0.92 \\
$_{74}$W$_{113}^{187}$ & 5.46 &0.83 & 13.2 & 0.23 & 0.51 \\
$_{75}$Re$_{113}^{188}$ & 5.73 &0.73 & 15.7 & 0.18 & 0.68 \\
\hline
\end{tabular}
\end{table}

The meaning and accuracy of the ``radiative" temperature, i.e., the one calculated
from the position of the maximum in the spectrum $\mathcal{E}(\eps)$; see formula
(21) and the table), will be easier to analyze if account is taken of certain
features of the thermodynamics of such a cooled body as a concrete nucleus. Owing
to the absence of fluctuations of its total energy $E$, the equilibrium temperature
of such a system becomes to a certain degree an approximate concept. The scale of
the related temperature uncertainty is given by the well known thermodynamic formula
(see [6])
\begin{equation}\label{27}
    \Delta T=T/\sqrt{C}
\end{equation}
for its fluctuation. The calculated values of $\Delta T$ are given in next to the
last column of the table.

The statements made above pertain to a definite nucleus with fixed composition.
One cannot exclude, e.g., the possibility that such limitations may become less
stringent when an attempt is made to ascribe a common temperature to an entire
aggregate of relatively close nuclei. It is therefore of interest to verify whether
an appreciable averaging of the radiative temperature takes place over the entire
atomic-weight interval $150 \leq A\leq 188$ where the comparison was made.
Using (15), we reduce all the temperatures to a single excitation energy, say
$E = 8$ MeV. We can then see that the swing of the fluctuations is in fair
agreement with the thermodynamic estimates of the variance $\Delta T$. Thus, the
characteristic period of the fluctuations along the $A$ axis is apparently small
in comparison with the region of nuclei under consideration, so that the averaging
referred to above indeed has time to occur
\footnote{This can also be confirmed qualitatively by means of theoretical estimates.
The variance $\Delta T$ of the temperature is closely related with the thermodynamic
fluctuation $\Delta S$ of the entropy of the closed system (see [6]). On the other
hand, in the particular case of an energy spectrum of the Fermi-liquid type,
the entropy, in order of magnitude, can be interpreted as the number of quasiparticles
that fall in the zone of the temperature smearing of the Fermi distribution. By
connecting the addition of not too large a number of nucleons to the nucleus,
on the one hand, with the possible change of the entropy of the quasi-particle,
on the other, we estimate the characteristic period of the fluctuations. It turns
out that about two such periods are subtended by the investigated interval
$150 \leq A \leq 188$. We do not present details of these qualitative estimates.}.
We are nevertheless left with the question of the systematic overestimate of the
temperature of the nucleus when the values of $\nu$ are too small (see above).

To highlight the distinction between the random (fluctuation) and systematic errors
more lucid, it is desirable to determine from experiment a quantity that
characterizes directly, if possible, the nuclear Fermi liquid as such. Satisfying
these requirements is the effective mass $m^*$; its value is also of definite
interest in itself. The effective mass determines the specific
heat and the entropy of the Fermi liquid [1,6]. The combinatorial expression for
the entropy reduces to an integral that can be calculated without difficulty
(the well-known problem of the specific heat of a degenerate Fermi gas reduces
to a similar procedure; see, e.g. [6]). Summing the contributions from the neutron
and proton quasiparticles, we have
\begin{equation}\label{28}
    S=\frac{4\pi}9\frac{R^2}{\hbar^2}m^*(\rho_f^N+\rho_f^Z)T.
\end{equation}

Just as in many other problems of nuclear physics, an important role is played
by the dimensionless variable
\begin{equation}\label{29}
    \rho_f=k_fR\gg 1,
\end{equation}
where $k_f$ is the limiting momentum of the quasiparticles of the corresponding
type and $R$ is the radius of the nucleus (more accurately, of equivalent volume).
When (28) is compared with (15), it is convenient to express the coefficient $a$
in the temperature dependence of the specific heat in terms of the energy and
the temperature. We then obtain for the effective mass
\begin{equation}\label{30}
    \frac{m^*}{m_n}=\frac{9/\pi}{\rho_f^N+\rho_f^Z}\frac{\hbar^2}{2m_nR^2}\frac{E}{T^2},
\end{equation}
where $m_n$ is the mass of the free nucleon. The question of the connection between
the ``limiting momentum" $\rho_f$ (see (29)) and the number of true particles in the
nucleus was considered earlier [9]. The corresponding formula is
\begin{equation}\label{31}
    N,Z=\frac4{9\pi}\rho_f^3-s\rho_f^2+q\rho_f.
\end{equation}

Since the relative accuracy of such an approximate expansion becomes better with
increasing number of nucleons, it is most natural to use data concerning the magic
numbers $82$ and $126$. This yields $s = 1.1$ and
$q = 6.8$ for the values of the parameters that enter in (31)
\footnote{Notice should be taken of the following: the expansion (31), and incidentally
the very concept of the radius $R$ of the nucleus, is of macroscopic accuracy.
Therefore the contributions of the lowest power of $\rho_f$ to the number of
particles (e.g., terms of order $\rho_f^0\sim 1$) are disregarded here. We obtained
initially the magic values $\rho_f$ (see [9], formula (19)), so that $s$ and $q$
were determined on the basis of data pertaining to spherical nuclei. It is easy to see,
however, that relation (31), with the same values of the parameters, remains valid
also for the case of non-spherical nuclei. Indeed, the first term in the right-hand
side is the quasiclassical limit for the number of cells in phase space; it depends
only on the volume of the system. The second (surface) term, on the other hand, can
generally speaking be influenced by the deformation. For equilibrium deformations of
non-spherical nuclei, however, we have $\alpha\sim\rho_f^{-1}$, and consequently
the relative change in the area of the surface of the nucleus is only a quantity
of the order of $\alpha^2 \sim\rho_f^{-2}$. This would yield, in final analysis,
a correction of the order of unity to the number of nucleons in the nucleus. Thus,
the actually realized equilibrium deformations exert no microscopic influence on
the form of the function $N(\rho_f)$, and the possible corrections would lie beyond
the limits of the accuracy of (31).}.

We assume
\begin{equation}\label{32}
    R=1.2\cdot 10^{-12}A^{1/3}\,\mathrm{cm}.
\end{equation}
The effective-mass values determined from (30) and (31) are given in the last column
of the table. The arithmetic mean is $m^*/m_n = 0.72$ (the point pertaining to the
nucleus Th$^{233}$ is also taken into consideration here; see footnote 7). This result,
however, was influenced by a systematic error, which lowers the effective mass of
the quasiparticle. Indeed, there is a striking correlation between the ratio $m^*/m_n$
calculated by this method and the number of quanta $\nu$ (see the table and the figure).
Consequently, a certain overestimate of the temperatures at extremely small $\nu$,
referred to above, indeed took place. In this case this source of error seems even
somewhat exaggerated, since the right-hand side of (30) contains the square of the
temperature. On the other hand, in the region $\nu = 3.5-4.1$, the criterion (26)
already seems to be fulfilled satisfactorily, as is confirmed also by the good
agreement over the spectra (see the figure). The best value is therefore probably
the one calculated for the six pertinent nuclei
\begin{equation}\label{33}
    m^*/m_n=0.87\pm 0.04
\end{equation}
(we give here the purely statistical mean-squared variance).

\section{Conclusions}

1.  The known treatment of the spectra of the evaporation neutrons by the
detailed-balance principle (see, e.g., [10]) makes it possible to carry out relatively
rough estimates of the nuclear temperature. The thermodynamics of electromagnetic
radiation of nuclei, which was developed in the present paper, is apparently more
quantitative in character. Therefore, given the corresponding experimental data,
it will be possible to measure more systematically and quantitatively the temperature
of other thermodynamic quantities at different excitation energies.

2.  At the present time, however, such a program could be realized only in part for
one of the regions of the non-spherical nuclei (see the preceding section). Another
interesting region begins with radium. Since many of the actinides in this region are
fissile, a study of the captured quanta should be carried out under conditions of
anticoincidences with fission fragments.

An even more noticeable shortcoming of the experimental data is due to their
limitation with respect to the excitation energy. At $E \sim 8$ MeV (the energy
of a thermal neutron), the conditions for the applicability of the theory are
frequently not very favorable, owing to the excessively low number of $\ga$ quanta in
the cascade (see the preceding section, and also (26) and footnote 5). Observation of
electromagnetic radiation of much more strongly excited nuclei is hindered by the
evaporation of neutrons. Therefore, for all their complexity, the performance of
corresponding experiments for anticoincidence with neutrons is extremely desirable.
They make it possible to overcome finally those difficulties in the reduction of the
experimental data, which we attempted to analyze in the preceding section\footnote{At
large $E$, the region where the criterion (25) is satisfied becomes much wider.
This makes it possible, in particular, to verify and refine the result (33) for the
effective mass of the nucleon (quasiparticle). We call attention to a possible more
profound shortcoming of the proposed method of measuring the effective mass, namely,
the right-hand side of (30) is inversely proportional to the square of the nuclear
radius $R$. It is clear that a transition to higher excitation energies does not
change the situation in this respect. The value (32) assumed above seems to agree
fairly well with the data on the scattering of electrons and on the internal
structure of the nuclei. However, the isotropic Fermi liquid is an object that is
rather exotic and is not frequently encountered in nature, so that the value
of the effective mass in a nuclear Fermi liquid can be of certain fundamental
interest. From this point of view, the question of the choice of the best value
of $R$ can still not be regarded as completely solved.}.
One can even hope that further development of the theory (see, in particular, the
next item), such
experiments will cast light on the interesting question of the phase transition
that takes place when a spherical nucleus is sufficiently excited.

3. Spherical nuclei owe their very existence to the residual interaction between
the quasiparticles. In [3] there were established only the most general, macroscopic
features of its structure. It may turn out, however, that the form of the
thermodynamic relations at low temperatures follows from it in a sufficiently
unique manner. After establishing the dependence of the specific heat on the
temperature (it is apparently not described by formula (15) in the given case),
one can attempt to develop also the theory of radiation of such nuclei.

Qualitative considerations give grounds for assuming that spherical nuclei are
characteristics by relatively high temperatures, and accordingly, by low entropies.
Therefore, in particular, the region most easily accessible to practice,
$E = 6-8$ MeV, calls for a critical review. If it turns out that at such
excitation energies the fluctuations are still large, then the thermodynamic
relations will probably be suitable here only for rough estimates.

We are grateful to A.I. Baz', V.V. Vladimirskii, I.I. Gurevich, M.V. Kazarnovskii,
A.A. Ogloblin, I.M. Pavlichenkov, V.P. Smilga, and K.A. Ter-Martirosyan for a
discussion of the results.

\bigskip
\bigskip

\centerline{\bf Figure caption}

\bigskip

The experimental spectra of the $\ga$-quantum cascades are shown by solid lines,
and the theoretical ones by dashed ones. The peaks in the soft parts of the
spectra are not of thermal origin; some of them could not be drawn at all in
the chosen scale.

\end{document}